\documentclass[runningheads]{llncs}

\usepackage{graphicx}
\usepackage{comment}
\usepackage{amsmath,amssymb}
\usepackage{color}
\usepackage{url}
\usepackage{hyperref}
\usepackage[square,sort,comma,numbers]{natbib}
\usepackage{textcomp}
\usepackage{xspace}
\usepackage{numprint}
\usepackage{booktabs}
\usepackage{subcaption}
\usepackage{cleveref}

\newif\ifreview
\reviewfalse

\ifreview
	\usepackage{lineno}

	\linenumbers
\fi

\begin{document}


\def\SubNumber{64}

\def\GCPRTrack{Main Track}

\title{Robust Tumor Segmentation with Hyperspectral Imaging and Graph Neural Networks}

\ifreview
\else
	\titlerunning{Robust Tumor Segmentation with HSI and GNNs}

	\author{Mayar Lotfy Mostafa\inst{1,2}\and Anna Alperovich\inst{1} \and
	 Tommaso Giannantonio\inst{1}\thanks{Corresponding author: tommaso.giannantonio@zeiss.com } \and Björn Barz\inst{1} \and Xiaohan Zhang\inst{3} \and  Felix Holm\inst{1,2} \and  Nassir Navab\inst{2} \and Felix Boehm\inst{4} \and Carolin Schwamborn\inst{4} \and Thomas K. Hoffmann\inst{4} \and Patrick J. Schuler\inst{4}
   }
	
	\authorrunning{M. L. Mostafa et al.}
	
	\institute{Carl Zeiss AG, Corporate Research \& Technology, Oberkochen, Germany \and CAMP, Technical University of Munich, Garching, Germany
	\and Carl Zeiss Meditec AG, Oberkochen, Germany \and
	 Dept. of Otorhinolaryngology, University Hospital Ulm, Ulm, Germany}
\fi

\maketitle              

\begin{abstract}
Segmenting the boundary between tumor and healthy tissue during surgical cancer resection poses a significant challenge.
In recent years, Hyperspectral Imaging (HSI) combined with Machine Learning (ML) has emerged as a promising solution.
However, due to the extensive information contained within the spectral domain, most ML approaches primarily classify individual HSI (super-)pixels, or tiles, without taking into account their spatial context.
In this paper, we propose an improved methodology that leverages the spatial context of tiles for more robust and smoother segmentation.
To address the irregular shapes 
of tiles, we utilize Graph Neural Networks (GNNs) to propagate context information across neighboring regions.
The features for each tile within the graph are extracted using a 
Convolutional Neural Network (CNN), which is trained simultaneously with the subsequent GNN.
Moreover, we incorporate local image quality metrics into the loss function to enhance the training procedure's robustness against low-quality regions in the training images.
We demonstrate the superiority of our proposed method using a clinical ex vivo dataset consisting of 51 HSI images from 30 patients.
Despite the limited dataset, the GNN-based model significantly outperforms context-agnostic approaches, accurately distinguishing between healthy and tumor tissues, even in images from previously unseen patients.
Furthermore, we show that our carefully designed loss function, accounting for local image quality, results in additional improvements.
Our findings demonstrate that context-aware GNN algorithms can robustly find tumor demarcations on HSI images, ultimately contributing to better surgery success and patient outcome.

\end{abstract}
\section{Introduction}

Head and Neck (H\&N) cancer, primarily squamous cell carcinoma (SCC), is the sixth most common cancer worldwide~\cite{parkin2005global} and is often lethal, with 350,000 cancer-related deaths~\cite{ferlay2015cancer,fidler2017assessing}.
Today, the primary treatment remains surgery, whose outcome is compromised by inadequate cancer margin detection in 10\% to 20\% of the cases \cite{marur2016head}. 
If residues of cancerous tissue remain, recurrence rates after surgery can be as high as 55\% \cite{marur2016head}.
In the typical surgical workflow, pre-operative imaging is used to plan tumor resection. Further, surgeons rely on experience and visual cues, as well as intra-operative pathologist consultations (IPCs) to provide guidance towards the total cancer resection. Yet, IPC is time consuming and may not fully reflect the delineation and extent of the lesion. 
Thus, tumor segmentation solutions based on optical imaging combined with machine learning recently emerged as an attractive non-invasive and real-time addition to the classical pathological testing~\cite{wang2021deep}.

\begin{figure}
        \centering
        \includegraphics[trim=0 90 0 20, clip, width=1\linewidth]{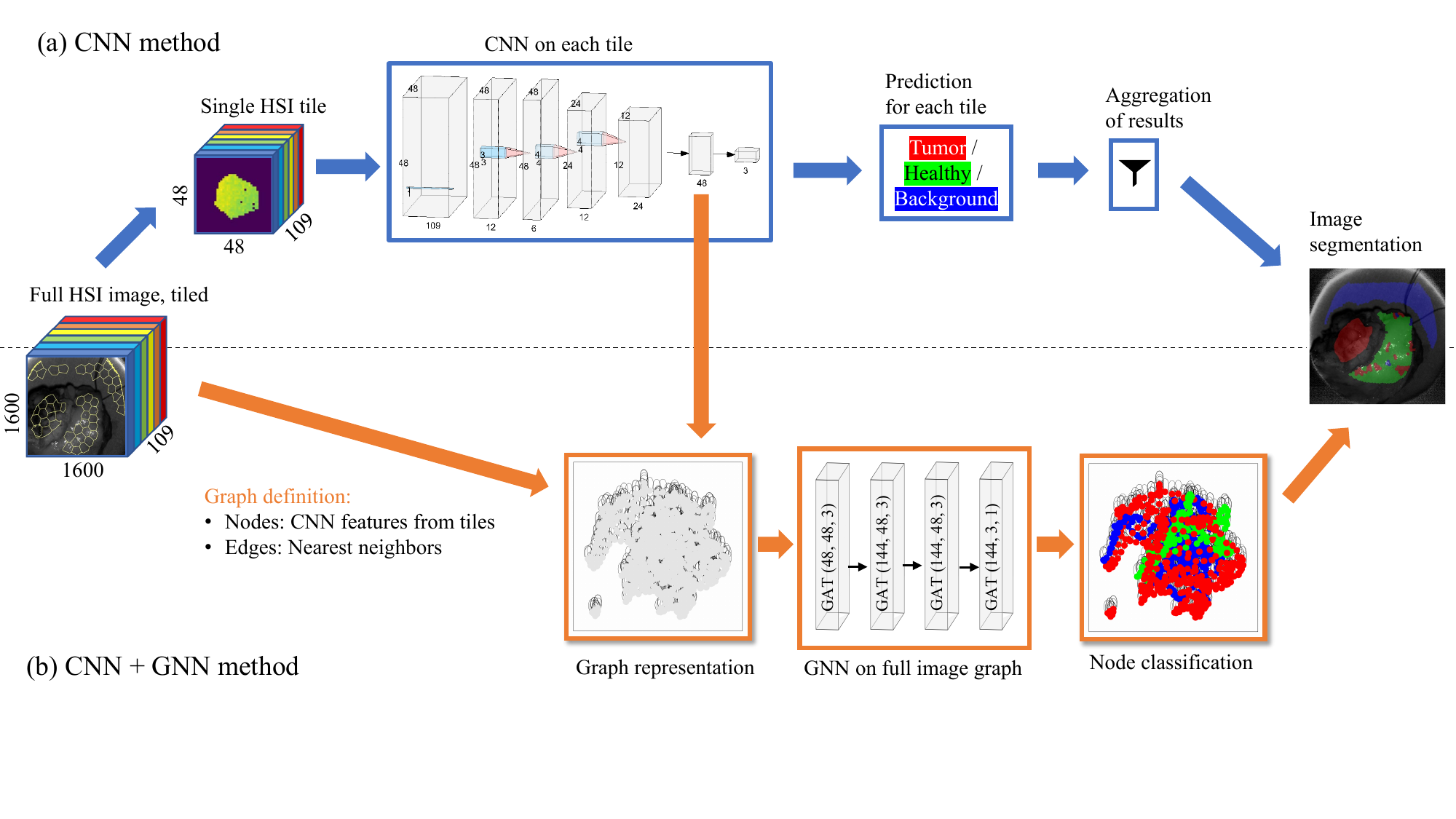} 
        \caption{Overview of the segmentation methods we consider. The top section (a, blue arrows) describes the baseline CNN method, while the bottom section (b, orange arrows) represents the CNN+GNN approach. Beginning from the left, we represent the full HSI data cube for one image (the grayscale shows luminosity at a single representative wavelength) with exemplary SLIC tiling overlaid in yellow. The numbers on the $x, y, \lambda$ axes represent the dimensions. Along the (a) path we show a representative cropped and zero-padded tile, the architecture of the CNN classifier model (each block represents a convolutional layer with feature dimensions  at the bottom, spatial dimensions at the left, and kernel dimensions near the blue rectangle), its possible predictions, and their aggregation to image segmentation. Along the (b) path, we show the graph representation of the full HSI image, incorporating as node features the CNN embeddings, the GNN model (each block represents a GAT layer with numbers of input and output channels and attention heads), and the resulting node predictions, which can be directly interpreted as image segmentation result.
        }
        \label{fig:segmentation_methods}
\end{figure}

Hyperspectral imaging (HSI) \cite{clancy2020surgical} is an
optical modality that combines the benefits of imaging and spectroscopy. It covers a continuous part of the visible/IR spectrum and images the whole field of view at hundreds of narrow wavelength bands.
HSI is an emerging technique in the biomedical laboratory and in the operating room \cite{lu2014medical,clancy2020surgical}, as it is non-invasive, non-ionizing, and label-free.
The information provided by the spectral properties of biological tissues can be used to distinguish them \cite{jacques2013optical}.
HSI has been used on biopsies, histopathological and fluorometric analysis, and disease biology~\cite{lu2014medical, fei2020hyperspectral, ortega2020hyperspectralREV}. 
In medical computer vision research, it was used for tumor tissue recognition of multiple cancer types \cite{halicek2019vivo, fabelo2019deep, giannantonio2023intra}.
These recent studies relied on the assumption that the spectral domain should contain most of the necessary information for tumor differentiation. Thus, single pixels or small patches are analyzed independently. 
This destroys any spatial context information, and the varying quality of the patches may bias the results.

\paragraph{Contributions} In this work, we present an ML method for overcoming these limitations.
By introducing context and application information to the model, we improve the prediction quality for the task of tumor/healthy tissue segmentation on unseen samples. Firstly, we introduce spatial context information by utilizing Graph Neural Networks (GNNs). We connect hyperspectral tiles based on their spatial proximity
into a graph and feed it to a GNN, whereby image segmentation is seen as a graph node classification task.
Our novel approach extracts tile-specific features with a CNN encoder and uses them as the node features for the GNN (see Fig.~\ref{fig:segmentation_methods}), combining spatial and spectra information. Secondly, we investigate whether the local quality of HSI data influences the segmentation results, finding that
tiles of a lower quality reduce performance.
We derive a custom loss function that weights the contribution from each spatial tile
based on custom quality metrics, thus improving the results further.
Our quantitative evaluations show that GNN models outperform CNN-based counterparts, improving accuracy. Qualitative results illustrate that predictions from GNN models are smoother. 
Furthermore, our quality-aware loss function further improves performance.

\section{Related work}
\label{sec:related_work}

\paragraph{Semantic segmentation of medical HSI data}

Different methods exist for for HSI-based tissue segmentation \cite{ortega2020information}, including 
physical \cite{leon2022hyperspectral} and data-driven models \cite{ghamisi2017advanced}, either classical \cite{moccia2018uncertainty, akbari2008wavelet} or based on deep learning \cite{li2019deep, khan2021trends, cui2022deep}.

Different strategies are possible for the combinations of spatial and spectral information. 
Single-pixel classification relies on spectral domain information
\cite{akbari2008wavelet}; spectral and spatial domains can otherwise be combined, e.g., by engineering spectral-spatial features \cite{he2017recent}. Most deep learning approaches have been based on multi-channel 2D CNNs, applied independently to macropixels (patches) to improve signal-to-noise and utilize some spatial information.
This was the strategy adopted by the HELICoiD project \cite{fabelo2016helicoid}, which achieved intra-operative brain tumor segmentation \cite{fabelo2019vivo} with a combination of classical and deep learning methods based on single pixels and patches \cite{fabelo2019deep,halicek2018tumor, manni2020hyperspectral, halicek2020tumor}; patch-wise fully 3D CNNs were also used \cite{halicek2018tumor, manni2020hyperspectral}. 
The same patch-based models were used for in vitro segmentation \cite{ortega2020hyperspectral2} and H\&N cancer (SCC) segmentation \cite{halicek2017deep, halicek2019hyperspectral, halicek2018tumor}.
A similar model was applied pixel-wise by \cite{urbanos2021supervised} on an independent brain cancer dataset,
and patch-wise using SLIC \cite{achanta2012slic} aggregation by \cite{giannantonio2023intra} (G23 henceforth) on low-grade glioma data. Full-image segmentation approaches were also proposed.
\cite{trajanovski2019tumor, trajanovski2020tongue} segmented an HSI dataset of tongue cancer using U-Net \cite{ronneberger2015u} based on patches.
\cite{cervantes2021automatic} segmented HSI data from hepatic and tyrhoid surgeries with U-Net on patches.
\cite{garifullin2018hyperspectral} segmented retinal HSI images using SegNet \cite{badrinarayanan2017segnet} and Dense-FCN \cite{jegou2017one}.
\cite{wang2020identification} defined a 3D Hyper-Net encoder-decoder model to segment HSI melanoma images. \cite{yun2021spectr} presented the Spectral Transformer model on a dataset of cholangiocarcinoma HSI images.
\cite{seidlitz2022robust} compared pixel- vs. patch-based vs. full-image organ segmentation on HSI animal images.

\paragraph{GNN segmentation of medical (non-HSI) data}

Graph Neural Networks (GNNs) \cite{wu2020comprehensive, chang2021comprehensive} are DL models that operate on data in graph form \cite{bondy1982graph}; they have been applied to diverse tasks in recent years, since the introduction of the seminal Graph Convolutional Network \cite{kipf2016semi}, GraphSAGE \cite{hamilton2017inductive} and Graph Attention Network (GAT) \cite{velivckovic2017graph}.
Any image can be seen as a graph, with the pixels as the nodes and the edges drawn to the neighbours. This is useful in case of HSI or volumetric data, which can be first aggregated into super-pixels (-voxels), and then processed as a graph with GNNs. The benefits are twofold: circumventing the cost of processing the raw data cube, and incorporating domain knowledge in the graph definition.
This approach has been applied to several medical image segmentation task.
\cite{yan2019brain} used GNNs to segment brain tissue from MRI data.
\cite{garcia2019joint} combined U-Nets with GNNs to segment airways from CT scans.
\cite{saueressig2021exploring, saueressig2021joint} performed brain tumor segmentation from MRI images.
\cite{duan2022semi} segmented eye fundus images with CNN and GNNs.
\cite{patel2023multi} segmented volumetric MRI data using GAT.
\cite{gaggion2022improving} defined the CNN+GNN encoder-decoder model HybridGNet for segmentation of chest X-ray data.
\cite{shao2022real} tracked liver tumor in X-ray data with GNNs and a biomechanical model.

\paragraph{GNN segmentation of non-medical HSI data}

Outside the medical domain, GNN models have been developed for HSI image segmentation, mostly for remote sensing and land coverage classification.
\cite{qin2018spectral} defined the spectral-spatial GCN that combines spectral and spatial similarity in the edge definition.
\cite{shahraki2018graph} combined a 1D CNN feature extractor with GCNs.
\cite{sha2019semi} employed edge convolutions, while \cite{hong2020graph} presented the miniGCN model.
\cite{wan2019multiscale} introduced the multi-scale dynamic GCN model; later from this model the context-aware dynamic GCN was derived
\cite{wan2020hyperspectral}.
Several variants were introduced \cite{liu2020semisupervised, mou2020nonlocal} also using GATs \cite{sha2020semisupervised, wang2020spectral} and GraphSAGE \cite{yang2020hyperspectral}.
\cite{zhang2021global} later introduced the Global Random Graph Convolution Network.
\cite{ding2023multi} proposed the multi-scale receptive fields graph attention network.
\cite{yao2023deep} presented a deep hybrid multi-graph neural network. 
To the best of our knowledge, no previous work used GNNs on medical HSI data.

\section{Methodology}
\label{sec:methods}

\subsection{Data}

\begin{figure}
            \centering
            \begin{subfigure}[b]{0.3\linewidth}
            \includegraphics[trim={0 28cm 0 10cm},clip,width=\linewidth]{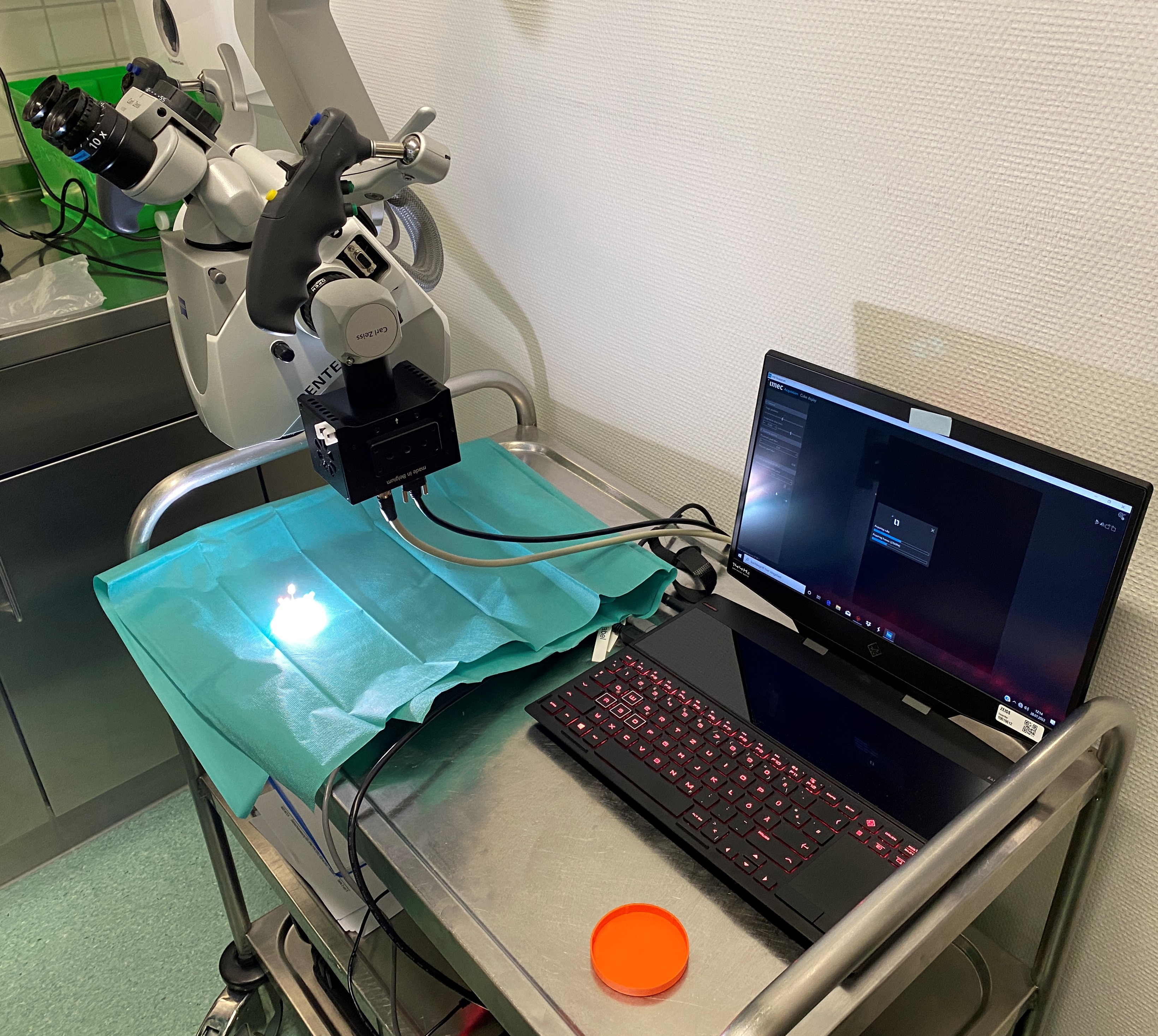}
            \caption{Laboratory setup}
            \end{subfigure}
            \begin{subfigure}[b]{0.3\linewidth}
            \includegraphics[width=\linewidth]{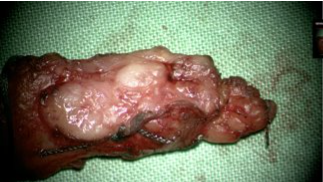} 
            \caption{RGB image}
            \end{subfigure}
            \begin{subfigure}[b]{0.3\linewidth}
            \includegraphics[width=\linewidth]{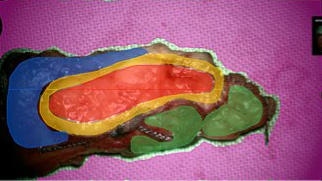} 
            \caption{Annotated image}
            \end{subfigure}
        \caption
        {\footnotesize Setup at the ENT department of University Hospital Ulm (a), RGB image of one exemplary resected tissue sample (b) and its corresponding labels (c), depicting in red, yellow, blue, green, and magenta the classes tumor, tumor margin, healthy, muscle and background respectively.} 
        \label{fig:setup}
\end{figure}

\paragraph{Data acquisition}

We follow the configuration decribed in G23; In brief, ex vivo HSI image acquisition was performed at the ENT department of University Hospital Ulm, employing an IMEC Snapscan VNIR hyperspectral camera in conjunction with a ZEISS OPMI\textsuperscript{®} PENTERO\textsuperscript{®} 900 surgical microscope through its optical port (Shown in Fig.~\ref{fig:setup}). 
The HSI setup shown in Fig.~\ref{fig:setup}(a) is prepared before the start of tumor surgery. Following tumor resection, the ex vivo tumor specimen undergoes a saline solution rinse to eliminate residual blood from its surface.
Next, the sample is taken to the laboratory-based HSI setup for data acquisition. 
For each specimen, a minimum of two HSI images are captured: an initial panoramic image covering the entire sample area to facilitate subsequent annotation of tissue categories, and an image with a reduced working distance and higher magnification to scrutinize the tumor center and margin in greater detail.
A consistent lighting scheme and region of interest are maintained by capturing an RGB photograph prior to each HSI image acquisition using the PENTERO microscope. The selection of microscope working distance and magnification values is determined based on the dimensions of the respective specimens. A constant light intensity of 50\% is used on the PENTERO microscope for all HSI measurements.
%
Post-operation, surgeons annotate diverse tissue classes within RGB images based on their prior knowledge from pre- and intra-surgical diagnostics.

Accurate annotation of ground truth is crucial in medical imaging studies to ensure reliable results, but the process of annotation is subjective, often performed by the surgeon who conducted the procedure. Despite this subjectivity, annotation benefits studies with limited duration. Our study considers the surgeon's experience and expertise, taking into account pathology assessment of the biopsies, to ensure accurate identification and delineation. While biopsy-only annotation would be ideal, it requires around 10 times more data points and years of clinical experiments. Furthermore, it poses a hardware limitation due to aging of light sources. Our approach balances accuracy and practicality, using both the surgeon's expertise and biopsy analysis, yielding reliable results despite hardware aging.

The RGB images and annotations are subsequently registered with their HSI counterparts.

\paragraph{Data characterization and preparation}
\label{sec:dataprep}

Our dataset comprises 49 SCC annotated images from 30 patients with resolution $1600 \times 1600$ pixels.
Every HSI image has $N_{\lambda} = 109$ wavelength channels between approximately 468~nm -- 790~nm.
Four classes were annotated by medical experts namely, tumor, tumor margin, healthy and muscle. We additionally annotated a background  class. 
For all experiments, we only consider the three classes tumor, healthy, and background.
We show in Fig.~\ref{fig:setup}(b,c) one RGB image and its annotations.
In order to boost signal-to-noise and support the segmentation task, we aggregate pixels into tiles (macro-pixels) by using a modified version of the SLIC algorithm \cite{achanta2012slic}, which aggregates pixels to minimize a combination of spatial and color distances between them. We set the color distance to be the Spectral Angle Mapper (SAM) distance, which for two pixels $i, j$ of spectra $\mathbf{s}_i, \mathbf{s}_j$ is given by $ \mathrm{SAM} (\mathbf{s}_i , \mathbf{s}_j) = \cos^{-1} (   (\mathbf{s}_i  \cdot  \mathbf{s}_j)/ (\lVert \mathbf{s}_i \rVert \,  \lVert \mathbf{s}_j \rVert  ) ) $.
Alternatively, the $L_2$ distance can be used, as given by $ L_2 (\mathbf{s}_i , \mathbf{s}_j) = (\sum_{l=1}^{N_{\lambda}}(s_{il} - s_{jl})^{2})^{1/2} \, $.
We set the number SLIC tiles to contain $\sim 200$ pixels each. We consider as possible proxies of each tile's $T_i$ quality its average pixel intensity $I_i$
(to isolate under-exposed and saturated regions) and the quantities $L_{2,i}$, $\mathrm{SAM}_i$ that are obtained by taking the average distances between the spectrum at each pixel $\mathbf{s}_{j}$ $\forall j \in T_i$  and the tile's average spectrum $\overline{\mathbf{s}}_i$, i.e., $ L_{2,i} \equiv \frac{1}{|T_i|} \sum_{j \in T_i} L_2(\mathbf{s}_{j}, \overline{\mathbf{s}}_i)$ (large spectral distances indicate low uniformity). (We drop the $i$ indices henceforth for simplicity.) 
The tile quality across an image varies. Therefore, we build a dataset with only high-quality tiles, following G23, and discard the tiles that have poor quality by applying threshold cuts on the three mentioned metrics: low spectral uniformity (we discard the upper 75\% percentile on both average SAM and $L_2$), extremely high or low mean pixels intensity (we average over channels and discard the lowest and highest 10\% percentiles in $I$), as well as tiles with non-uniform class labels.
This results in a dataset consisting of high-quality tiles only. The total amount of tiles in our dataset is \numprint{36693}, of which \numprint{26190} are high-quality tiles.
We split the data into training+validation and testing samples patient-wise, retaining all images from six patients for testing, i.e., we use 39 images for training and validation and 10 images for testing.
We divide accordingly both full and high-quality dataset into three sets containing 
65\%, 16.5\%, and 18.5\% of the
tiles for the train, val, test respectively.

\subsection{Tissue segmentation methods}
\label{sec:NN_methods}

We consider the two main segmentation methods shown in \cref{fig:segmentation_methods}, CNN-based and CNN+GNN-based, both of which can be extended with our quality-weighted loss function.

\paragraph{CNN method}

The CNN-based method closely follows the approach of Ref.~\cite{giannantonio2023intra}.
Here we isolate the hyperspectral SLIC tiles and spatially zero-pad them to a common shape of $48 \times 48 \times 109$ ($x \times y \times \lambda$),
so that they can be passed to a CNN model for classification. We then aggregate the CNN predictions into a composite image segmentation result.

The CNN model uses a similar encoder architecture to Ref.~\cite{giannantonio2023intra}.  We first apply a channel compression layer, implemented with a $1  \times 1$ convolutional kernel with stride 1 and no padding, to reduce the number of spectral dimensions from $N_{\lambda} = 109$ to $N_{\mathrm{ch}} = 12$. We then apply four 2D convolutional blocks bringing the number of features from $N_{\mathrm{ch}}$ to $(N_f / 2, N_f, 2 N_f, 4 N_f)$ respectively, where we take $N_f = 12$. The kernel sizes are respectively $(3, 4, 4, 4)$, the strides $(1, 2, 2, 2)$ and padding $(1, 1, 1, 1)$. 
The resulting features of dimensions $48 \times 6 \times 6$ are then average-pooled, thus yielding for each tile $i$ an encoded representation $\mathbf{z}_i \in \mathbb{R}^{48}$, which we pass to a fully connected classification head. 
We apply a LeakyReLU activation and BatchNorm after each convolutional layer.

During training we apply the following data augmentation: a 50\% dropout before the classification layer, a random shift along spatial axes, a random pixel brightness shift, and random rotation, rescaling, and Gaussian blurring.

Our baseline strategy consists of training only on the high-quality tiles (training and validation splits) obtained with the procedure described in \cref{sec:dataprep}, using a regular cross-entropy loss function summed over all tiles. At test time, we always evaluate the model on all annotated image areas (test split, whereby tiles are generated on the test data on the fly, so the exact number will vary).

\paragraph{CNN with quality-weighted loss}

While the exclusion of low-quality tiles at train time generally makes the models more robust on the good-quality areas, in any realistic surgical setting it would be desirable to have meaningful predictions over the full image, including lower-quality areas, such as blood vessels and border regions.

\begin{figure}
    \centering
    \includegraphics[width=0.55\linewidth]{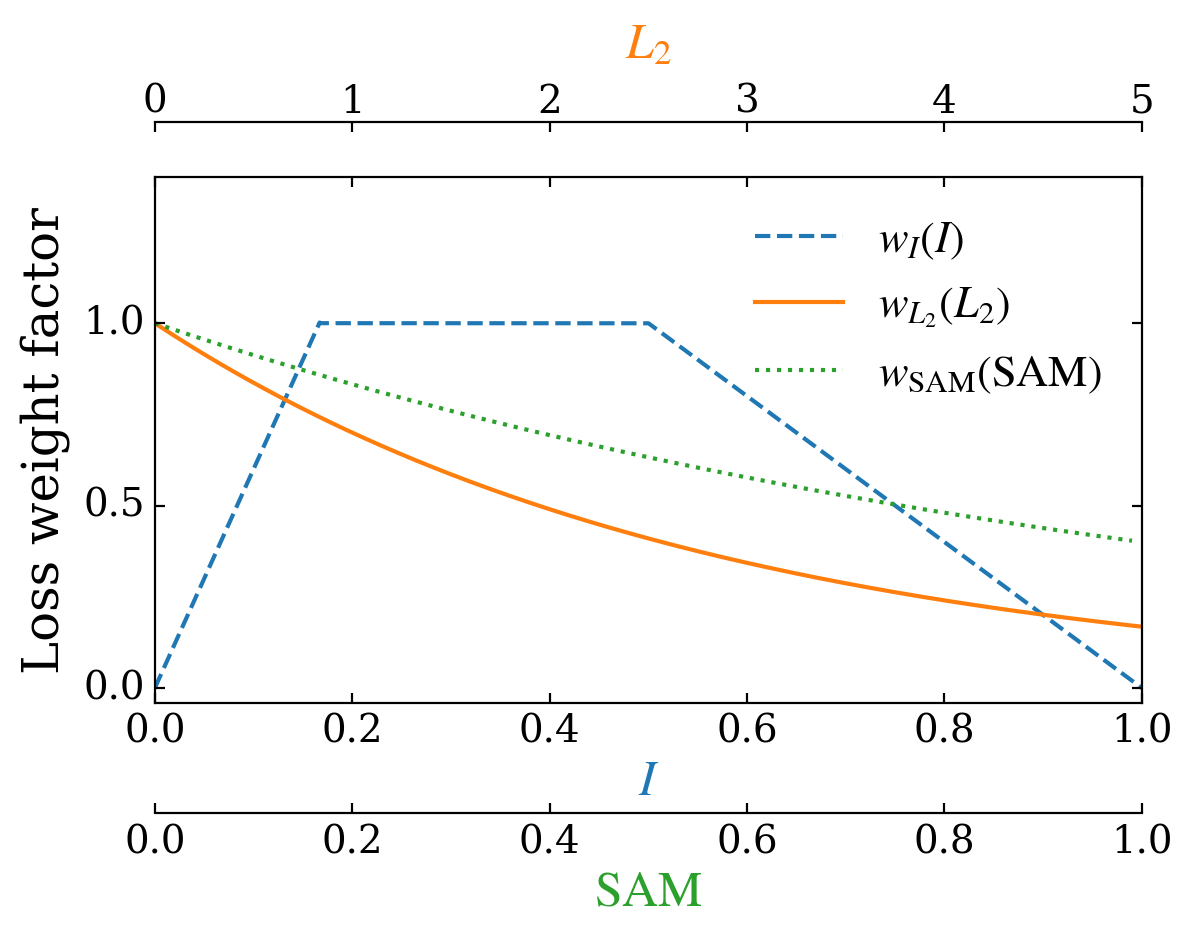}
    \vspace*{-4mm}
    \caption{Loss weight factors depending on tile quality metrics defined in Eq.~(\ref{eq:weights}). 
    }
    \label{fig:weight_factor}
\end{figure}

To address this requirement, we incorporate all tiles to the training and validation sets while introducing a quality-weighted loss function to downweight low-quality tiles. We define the loss weighting factor $w_i$ for each tile $T_i$ as
\begin{equation}
    w_i = w_I (I_i) \cdot w_{L_2} (L_{2,i}) \cdot w_{\mathrm{SAM}} (\mathrm{SAM}_i) \, ,
\end{equation}
where the quality metrics $I$, $L_{2}$, $\mathrm{SAM}$ are defined in \cref{sec:dataprep}.
Based on visual inspection of HSI images, we choose for the three components the forms that reflect the quality of the images (see Fig.~\ref{fig:weight_factor}):
\begin{eqnarray}
    w_I (I) &=& 
    \begin{cases}
    6 I & \text{if} \:\:\:\: I \leq  0.167  \\ 
    1 & \text{if} \:\:\:\: 0.167 < I < 0.5 \\ 
    2 - 2 I & \text{if} \:\:\:\: I \geq  0.5 
    \end{cases} \nonumber \\
    w_{L_2}(L_{2}) &=& 0.7^{L_{2}} \nonumber \\
    w_{\mathrm{SAM}}(\mathrm{SAM}) &=& 0.4^{\mathrm{SAM}}\, .
\label{eq:weights}
\end{eqnarray}

With this strategy, the model during training also uses the low-quality tiles, albeit with less weight, so that information from those regions is not lost. 
At test time, we expect better generalization, especially in the regions with high density of blood vessels, low light, and close to saturated areas.

We compare the results obtained by training on all tiles with the weighted loss with the model trained on good tiles only, as well as with the same model trained on all tiles with a regular cross-entropy loss.

\paragraph{CNN + GNN method}
We improve over the CNN baseline by incorporating the spatial context of the tiles using GNNs. 
For every HSI image, we construct an undirected graph $\mathcal{G} = (\mathcal{V}, \mathcal{E})$ with vertices $\mathcal{V}$ and edges $\mathcal{E}$. Here $N = |\mathcal{V}|$ is the number of nodes, and $A \in \mathbb{R}^{N \times N}$ is the adjacency matrix. In this context, the vertices represent the tiles, and the edges their relational structure.

In order to characterize the nodes, we need to aggregate the tiles' spectral information into a matrix $ X \in \mathbb{R}^{N \times F}$, where $F$ is the dimensionality of the aggregated features per tile. 
To encode intra-tile structure in both spatial and spectral domains, we choose to use
our baseline CNN model as a feature extractor, thus setting $X = Z \in \mathbb{R}^{N \times 48}$,
where we aggregate the latent space representation of all tiles into one matrix $Z$.
Differently from the baseline CNN model, in this case we do not apply batch normalization to the CNN, and we use for the final dropout layer of the CNN model a 30\% probability, as the GNN is less prone to overfitting.

We define the adjacency matrix $A$ by simply selecting for each tile its $k$ nearest neighbors with a kNN algorithm, using $k = 2$; we find this small number of neighbors to be effective in combating overfitting, which we observed in increasing amounts for $k \in \{3, 5\}$.

The graphs thus defined are then passed to a Graph Attention Network (GAT) \cite{velivckovic2017graph} with 64 hidden channels, 3 hidden GAT layer and 3 attention heads. After each GAT layer, a ReLU activation is applied, and before the last layer we apply a 30\% dropout during training.
While we experimented with different GNN choices, we found the GAT model to be suitable for our use-case as opposed to simpler graph-convolutional layers such as the GCN \cite{kipf2016semi}.
The attention mechanism provides the flexibility of applying different weights for different nodes in each tile's neighborhood, which is likely to be beneficial, e.g., near tissue class boundary regions. We note however that a full optimization of the GNN architecture is beyond the scope of this work.

During training, in addition to the CNN-level augmentation described above, we apply graph-level data augmentation by randomly jittering the spatial coordinates associated with each and by randomly dropping up to 30\% of the nodes.
We use per-node cross-entropy as the loss function. 
While it is possible to train the CNN+GNN model end-to-end in principle, we found that in order to reduce overfitting, it is beneficial to not let gradients flow from the GNN head into the CNN backbone during backpropagation, and instead minimize the combined loss $L_{\mathrm{comb}} = L_{\mathrm{CNN}} + L_{\mathrm{GNN}}$, where both losses are cross-entropy.

As for the CNN case, our baseline strategy is to first train the model using graphs constructed from the good-quality tiles only; at test time we always evaluate the model on all annotated image areas of the 10 left-out images.
Then we include all tiles in the training  by introducing the quality-weighted loss.
Also here, we will compare the results obtained by training on graphs constructed from all tiles using weighted and unweighted losses with the baseline model trained on graphs constructed from good tiles only.

\section{Results}
\label{sec:results}

We trained all models described in \cref{sec:NN_methods} on a Linux machine with 12 CPU cores, 64 GB RAM, and an NVIDIA GeForce RTX\texttrademark\ 3090 GPU (24 GB RAM).

\subsection{Quantitative evaluation}

\begin{figure}[h]
            \centering
            \includegraphics[width=0.55\linewidth]{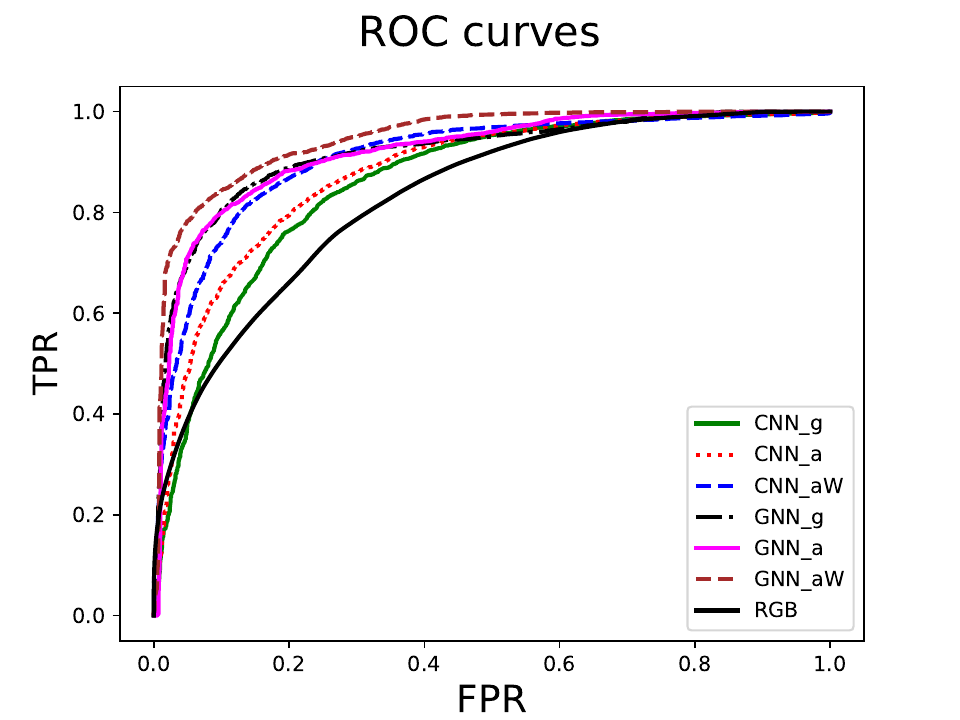}
        \caption
        {\footnotesize We compare the performance of all approaches by visualizing ROC curves. The plot illustrates that the baseline CNN method that was trained on only good-quality tiles performs worse than the other methods. All GNN approaches outperform pure CNNs and GNN with quality-aware backbone achieve the best performance. } 
        \label{fig:roc}
\end{figure}

We summarize in Tab.~\ref{table:models_comparisons} the quantitative results of all experiments evaluated on the test set that includes all annotated image areas.
Here we report the accuracy, $F_1$-score and IoU~\cite{muller2022towards} 
metrics per class (H, T, B for healthy, tumor, and background respectively) and macro-averaged across all classes. Additionally, we report the integrated area under the ROC curve (AUC) for the tumor/healthy binary classification problem, where we excluded the background class, since it is irrelevant for the clinical application.
The ROC curves for all models are shown in Fig.~\ref{fig:roc}.

In the analysis of the results, the closest previous work we can compare to is the method by G23, which is equivalent to our CNN\_g model;
other CNN-based models build on top of this.
While the CNN models already reach nontrivial accuracy, surpassing 80\% in most cases, the GNN method leads to an overall superior performance. All best results of Tab.~\ref{table:models_comparisons} lie in the GNN section, and a like-to-like comparison shows that the GNN approach improves average accuracy by 11 percent points (pp) for training on good-quality tiles compared to the CNN\_g baseline, by 1 pp for all tiles, and by 2 pp in the weighted loss case.

Models trained on all tiles without weighted loss function yield contradicting results: they show a better performance on the healthy class, but in some cases yield degraded results on the tumor class. 
This can be explained by the decreased quality of the dataset, where individual tiles contain heterogeneous spectra and suffer from high noise and signal saturation. 
The introduction of a weighted loss improves the results compared to the use of all tiles without weights, by rebalancing the performance between classes. The healthy class quality decreases, while the tumor class quality increases, leading to an average accuracy improvement of 2 pp (3 pp) for the CNN (GNN) methods.
Quantitative results according to the different metrics are largely consistent: GNN methods are superior, and the usage of a weighted loss improves most metrics. The AUC provides a good single-number summary of each model quality for clinical applications, see~\ref{fig:roc}. We observe once again that the GNN models are superior, the addition of a weighted loss is beneficial, and each GNN model is ahead of its CNN counterpart.
Our best model, GNN\_aW, is according to this metric 9 pp ahead of the baseline CNN\_g.

\begin{table*}
\centering
\caption{\footnotesize Quantitative comparison of model performance. We report accuracy, $F_1$-score, and IoU on each class (H, T, B refer to healthy, tumor and background respectively) and macro-averaged across all classes (Avg); we also report the ROC AUC for healthy/tumor binary classification. Models trained on good tiles only, on all tiles, and on all tiles with a quality-weighted loss are denoted with \_g, \_a, \_aW respectively. Best results are in bold. The CNN\_g model is equivalent to G23.}
\resizebox{0.9\linewidth}{!}{
\begin{small}
\begin{tabular}{c cccc cccc cccc c }
\toprule
            & \multicolumn{4}{c}{Accuracy} & \multicolumn{4}{c}{$F_1$-score} & \multicolumn{4}{c}{IoU} & AUC \\ 
Model       & H & T & B & \textit{Avg.} & H & T & B & \textit{Avg.} & H & T & B & \textit{Avg.} & H vs. T \\ 
\midrule
CNN\_g (G23) & 0.83 & 0.68 & 0.86 & \textit{0.79} & 0.79 & 0.63 & 0.83 & \textit{0.75} & 0.68 & 0.48 & 0.75 & \textit{0.64} & 0.86 \\ 
CNN\_a  & 0.85 & 0.76 & 0.94 & \textit{0.85} & 0.82 & 0.69 & 0.92 & \textit{0.81} & 0.71 & 0.56 & 0.71 & \textit{0.66} & 0.88 \\ 
CNN\_aW & 0.84 & 0.80 & 0.97 & \textit{0.87} & 0.85 & 0.71 & 0.88 & \textit{0.81} & 0.74 & 0.58 & 0.73 & \textit{0.68} & 0.91 \\ 
\midrule
GNN\_g  & 0.78 & \textbf{0.92} & \textbf{1.00} & \textbf{\textit{0.90}}  & 0.84 & \textbf{0.77} & 0.91 & \textit{0.84} & 0.73 & \textbf{0.65} & 0.78 & \textit{0.72} & 0.92 \\ 
GNN\_a  & \textbf{0.94} & 0.64 & 0.99 & \textit{0.86} & 0.85 & 0.67 & \textbf{0.99} & \textit{0.84} & 0.75 & 0.50 & \textbf{0.84} & \textit{0.69} & 0.92 \\ 
GNN\_aW & 0.87 & 0.83 & 0.98 & \textit{0.89} & \textbf{0.87} & 0.74 & 0.95 & \textbf{\textit{0.85}} & \textbf{0.78} & 0.60 & 0.82 & \textbf{\textit{0.73}} & \textbf{0.95} \\ 
\midrule
RGB  & 0.89 & 0.47 & 1.0 & 0.78 & 0.35 & 0.51 & 0.66 & 0.50 & 0.44 & 0.36 & 0.50 & 0.44 & 0.83 \\ 
\bottomrule
\end{tabular}
\end{small}
}
\label{table:models_comparisons}
\end{table*}

\begin{figure}[h!]
    \centering
    \begin{minipage}{0.14\textwidth}        
            \centering
            \includegraphics[width=\textwidth]{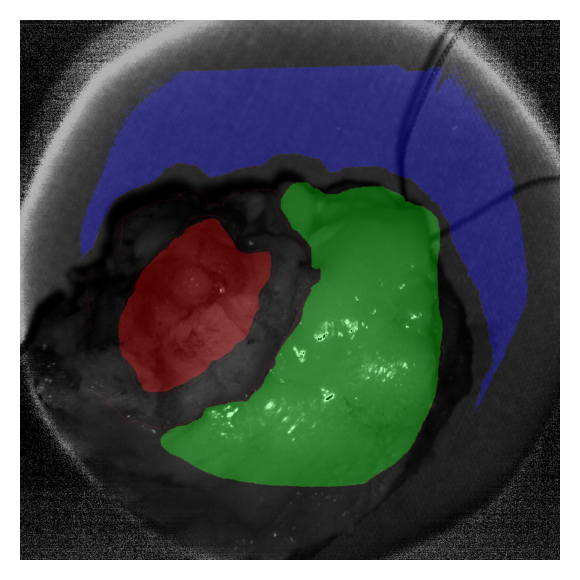}
    \end{minipage}%
    \begin{minipage}{0.14\textwidth}        
            \centering
            \includegraphics[width=\textwidth]{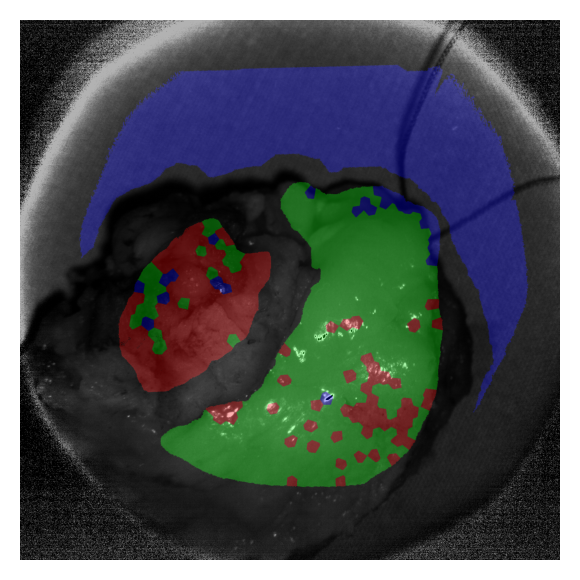}
    \end{minipage}%
    \begin{minipage}{0.14\textwidth}        
            \centering
            \includegraphics[width=\textwidth]{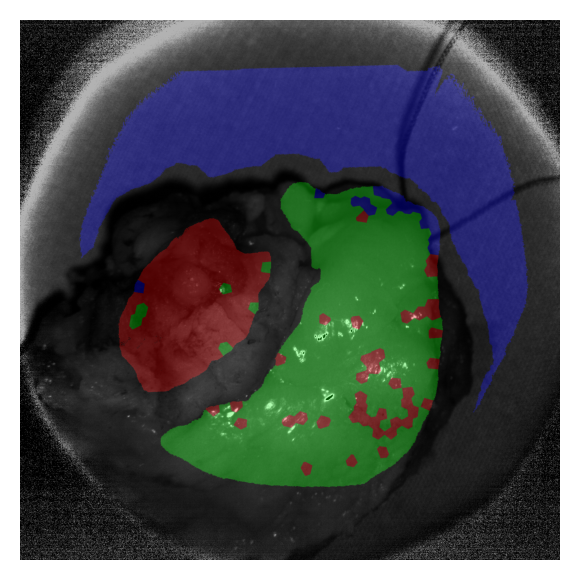}
    \end{minipage}%
    \begin{minipage}{0.14\textwidth}        
            \centering
            \includegraphics[width=\textwidth]{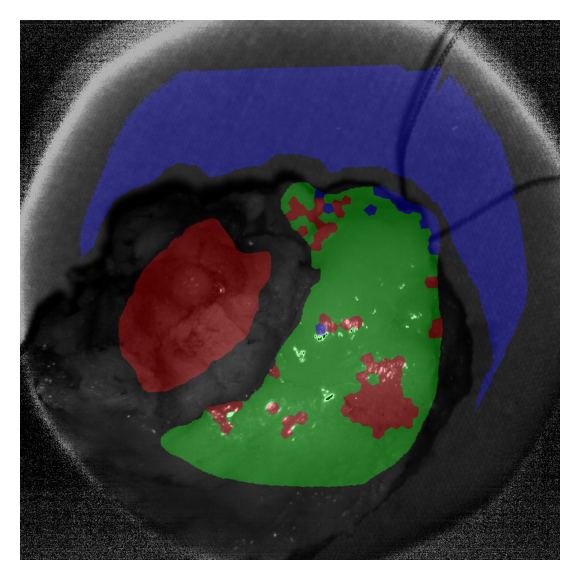}
    \end{minipage}%
    \begin{minipage}{0.14\textwidth}        
            \centering
            \includegraphics[width=\textwidth]{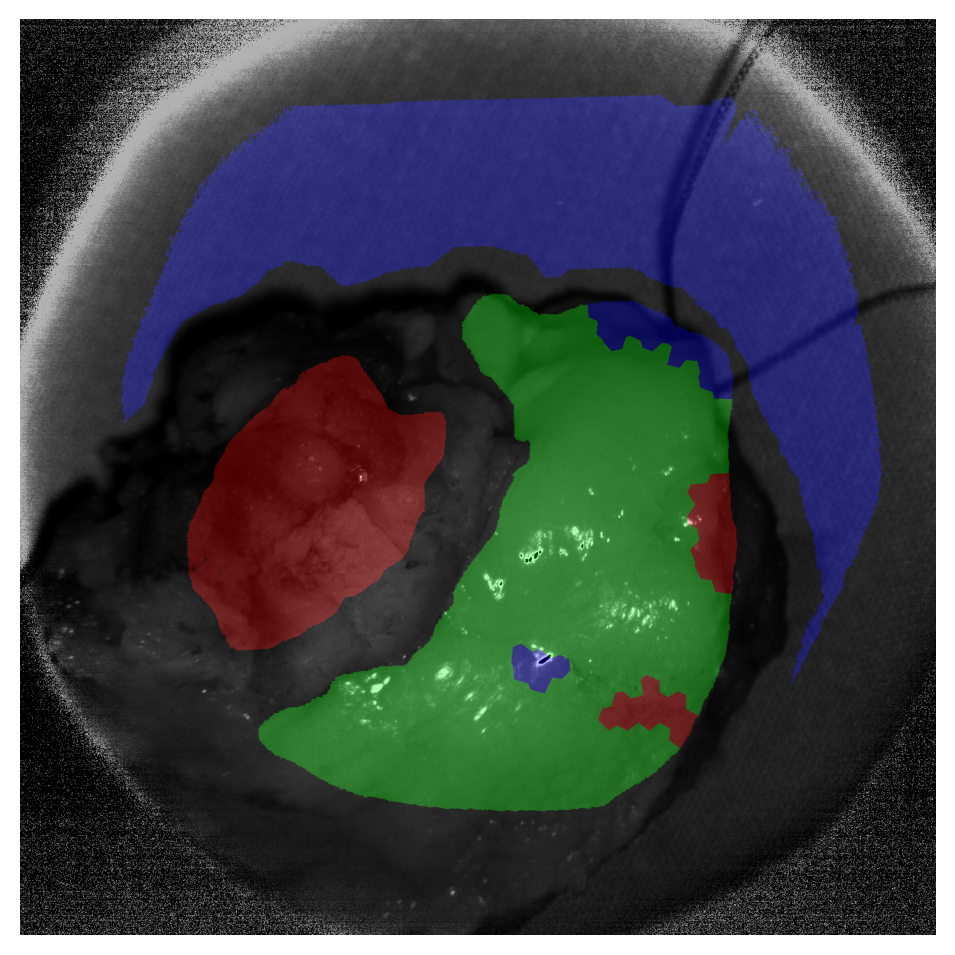}
    \end{minipage}%
    \begin{minipage}{0.14\textwidth}        
            \centering
            \includegraphics[width=\textwidth]{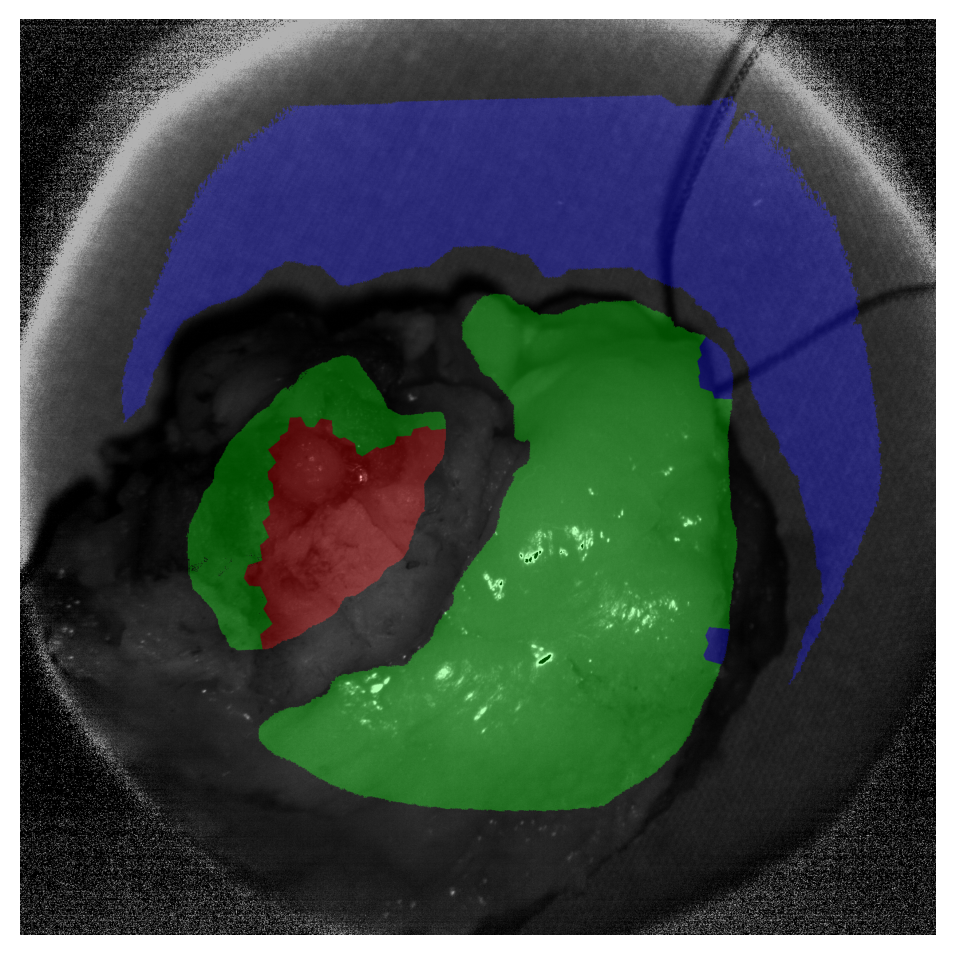}
    \end{minipage}%
    \begin{minipage}{0.14\textwidth}        
            \centering
            \includegraphics[width=\textwidth]{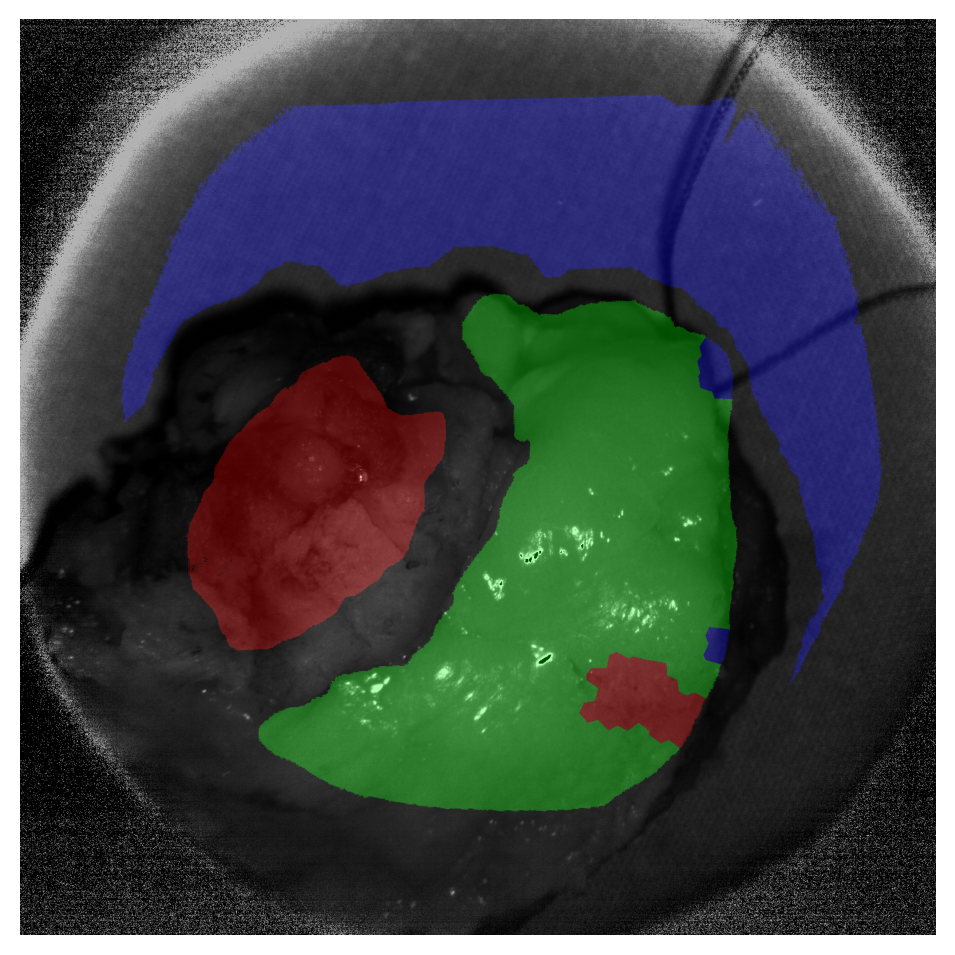}
    \end{minipage}
          \newline
    \begin{minipage}{0.14\textwidth}        
            \centering
            \includegraphics[width=\textwidth]{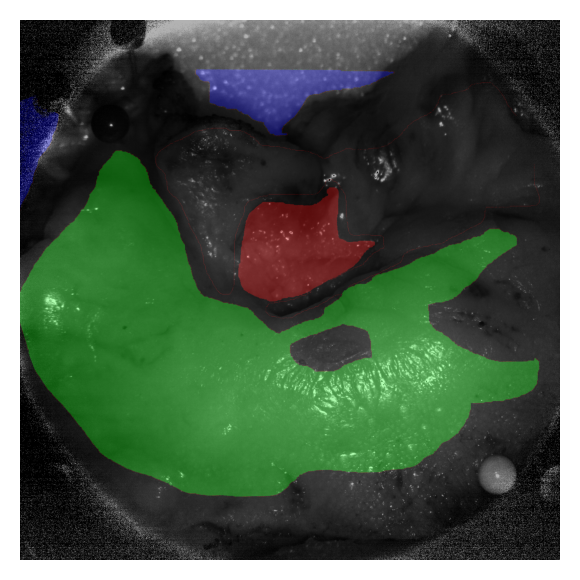}
    \end{minipage}%
    \begin{minipage}{0.14\textwidth}        
            \centering
            \includegraphics[width=\textwidth]{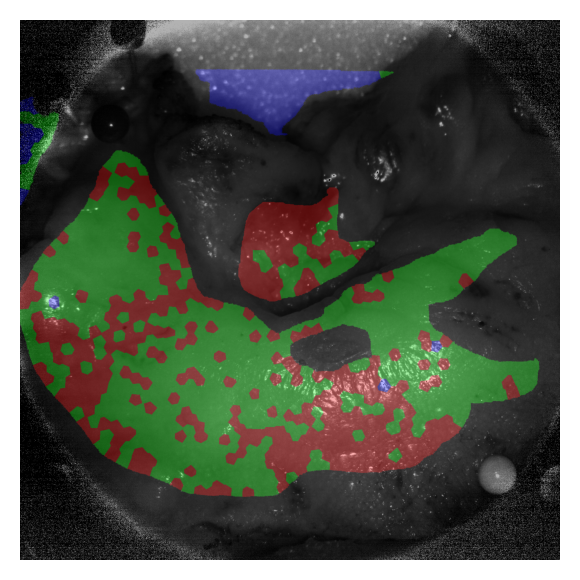}
    \end{minipage}%
    \begin{minipage}{0.14\textwidth}        
            \centering
            \includegraphics[width=\textwidth]{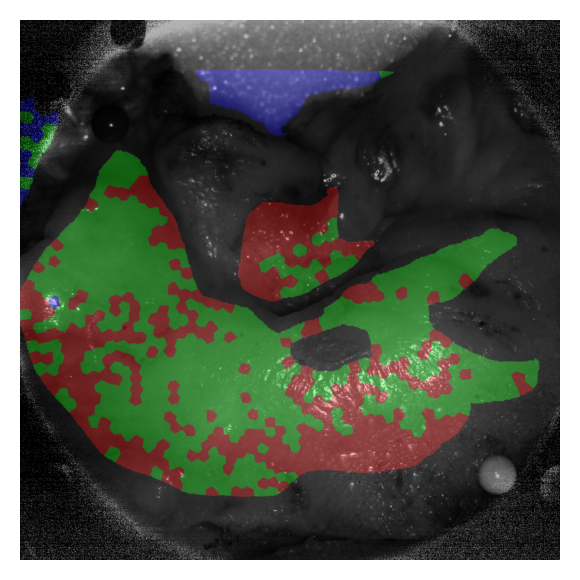}
    \end{minipage}%
    \begin{minipage}{0.14\textwidth}        
            \centering            
            \includegraphics[width=\textwidth]{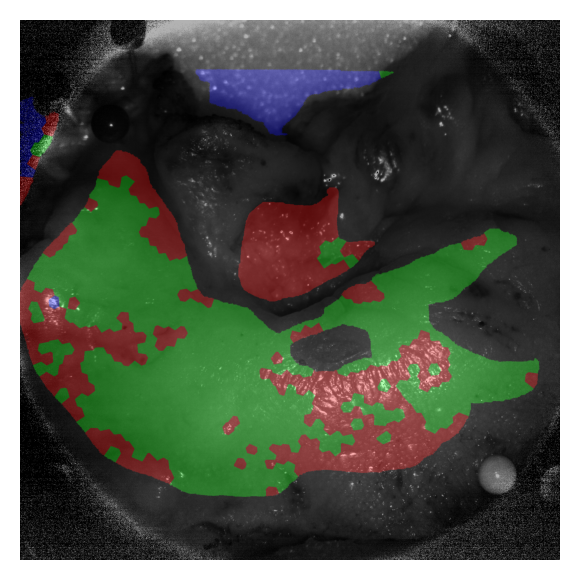}
    \end{minipage}%
    \begin{minipage}{0.14\textwidth}        
            \centering
            \includegraphics[width=\textwidth]{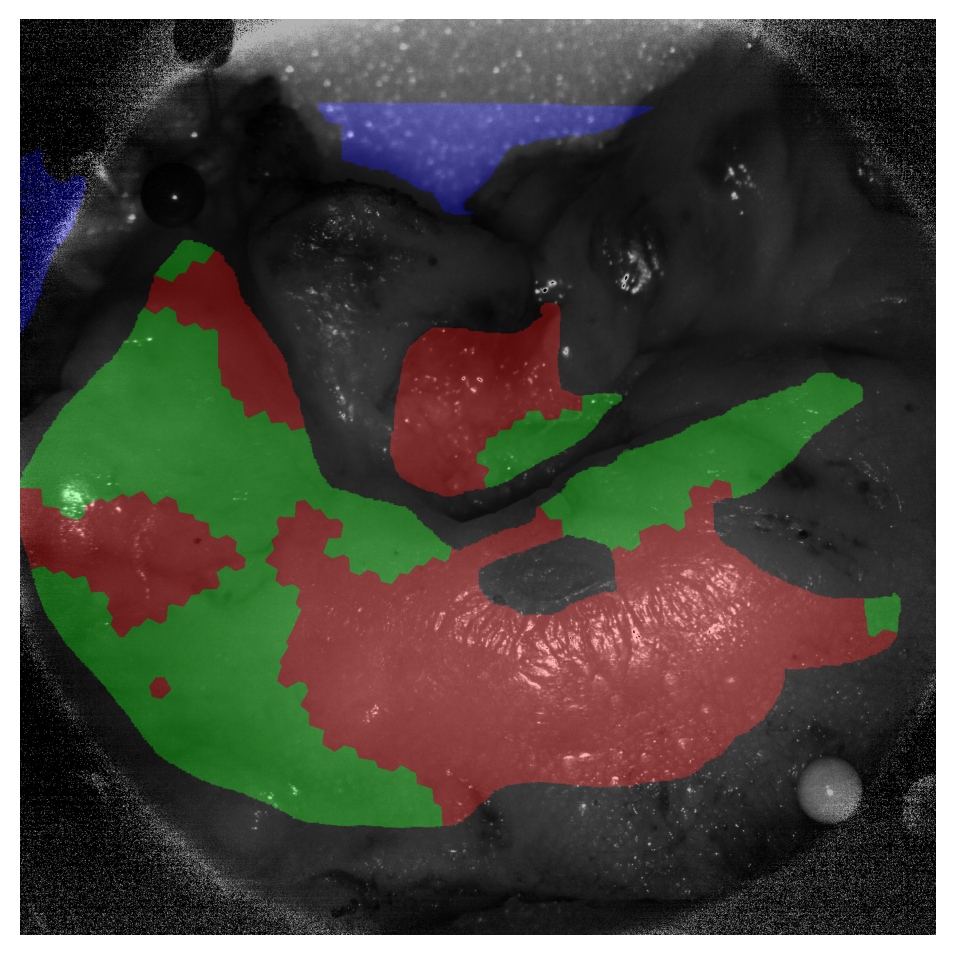}
    \end{minipage}%
    \begin{minipage}{0.14\textwidth}        
            \centering
            \includegraphics[width=\textwidth]{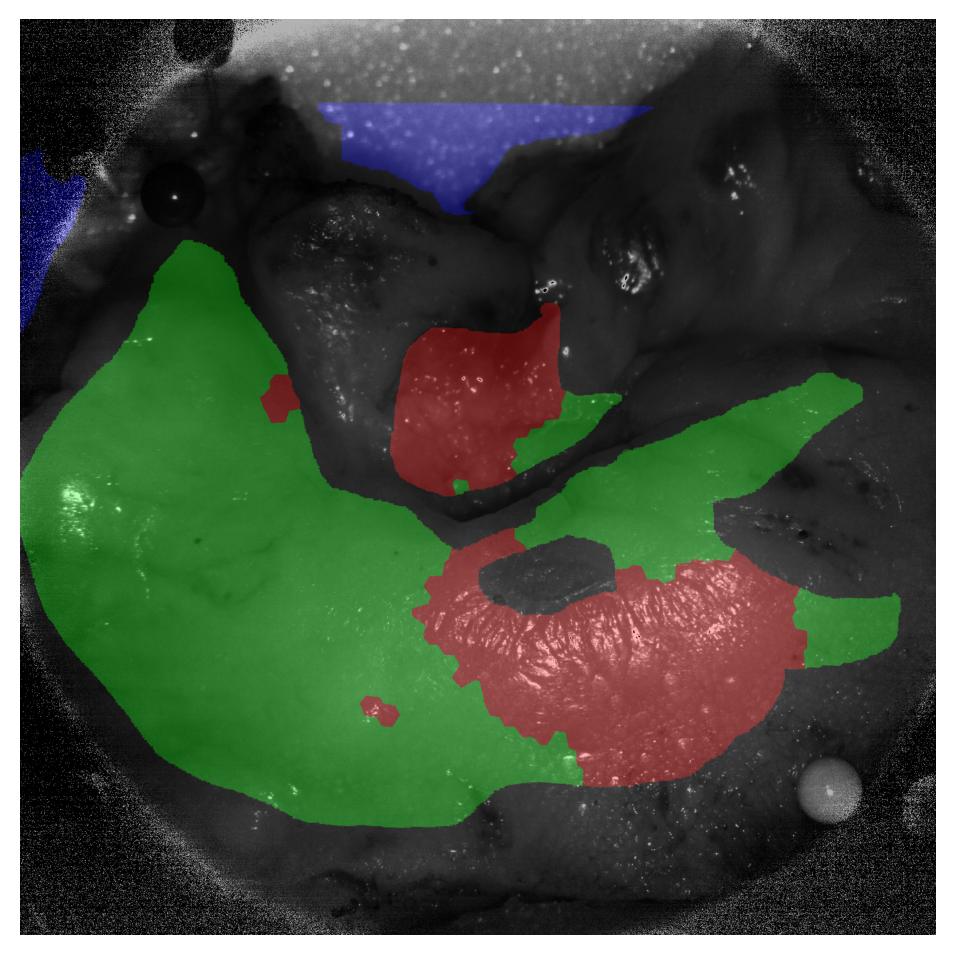}
    \end{minipage}%
    \begin{minipage}{0.14\textwidth}        
            \centering
            \includegraphics[width=\textwidth]{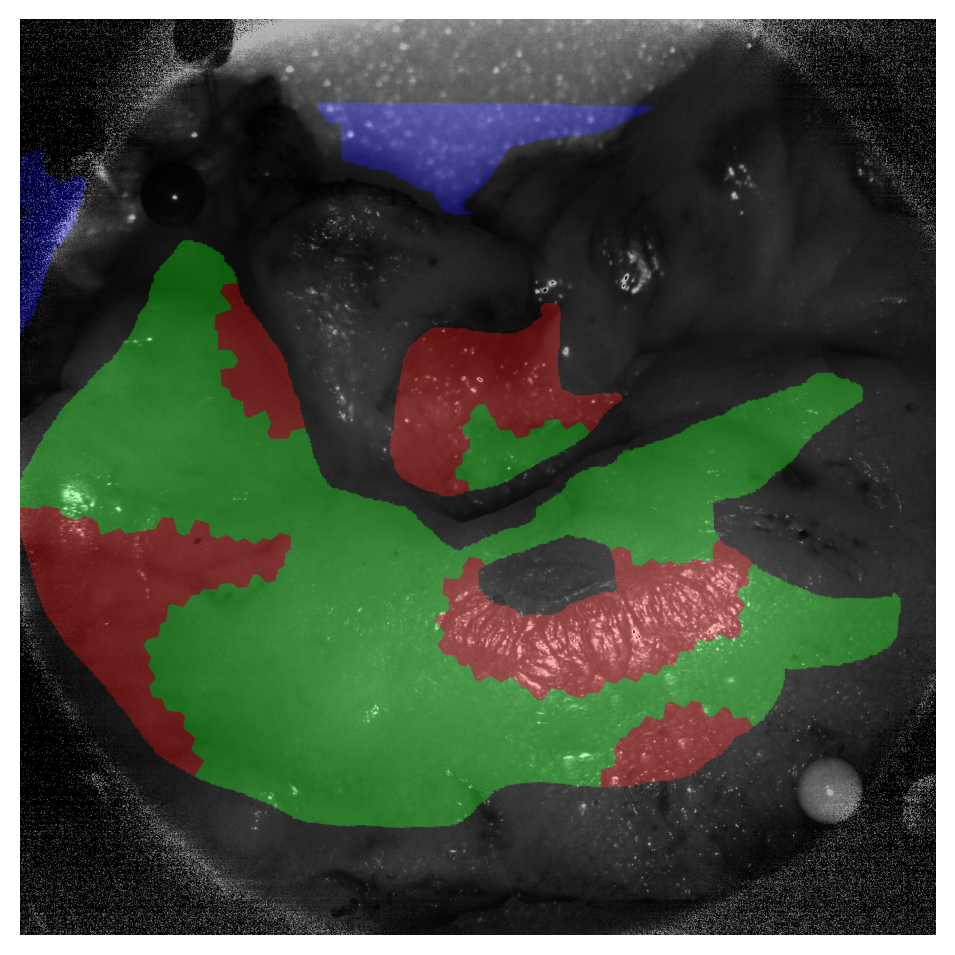} 
    \end{minipage}
        \newline
    \begin{minipage}{0.14\textwidth}        
        \centering
            \includegraphics[width=\textwidth]{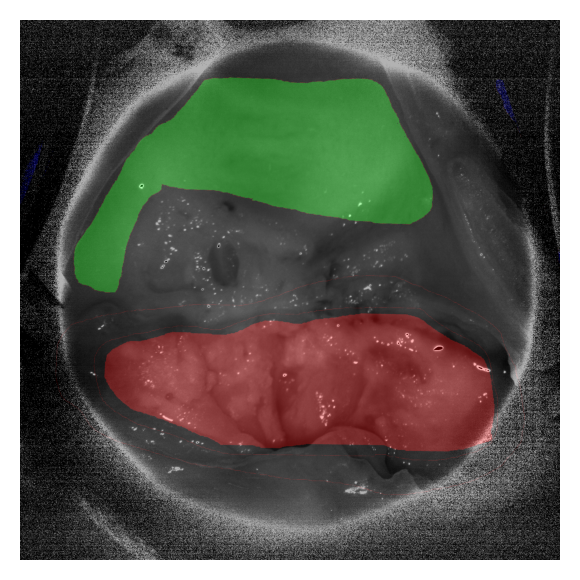}
    \end{minipage}%
    \begin{minipage}{0.14\textwidth}        
        \centering
        \includegraphics[width=\textwidth]{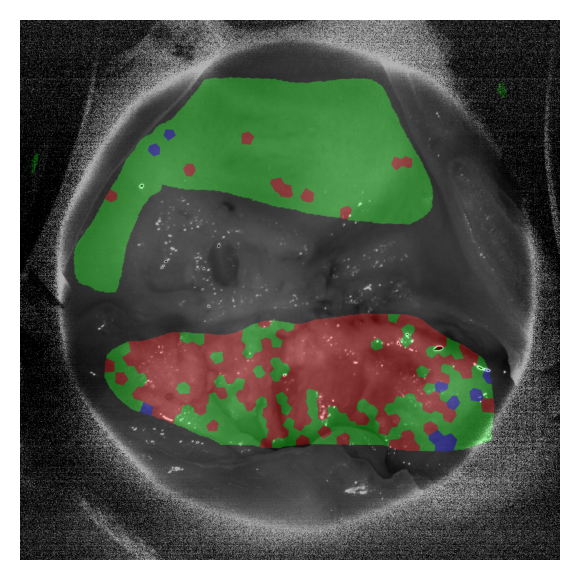}
    \end{minipage}%
    \begin{minipage}{0.14\textwidth}        
            \centering
            \includegraphics[width=\textwidth]{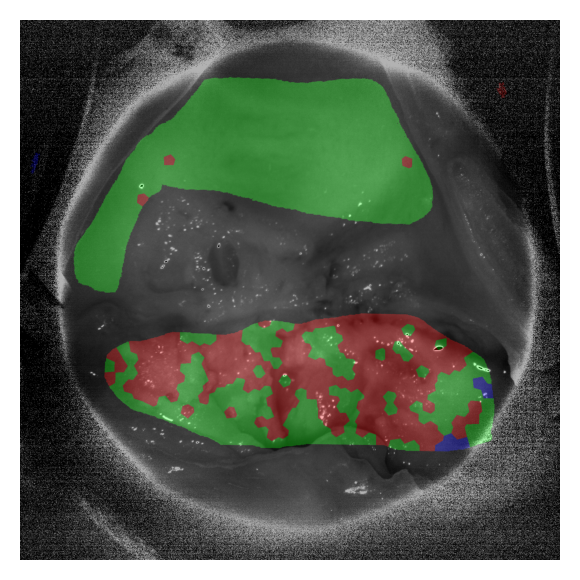}
    \end{minipage}%
    \begin{minipage}{0.14\textwidth}        
            \centering
            \includegraphics[width=\textwidth]{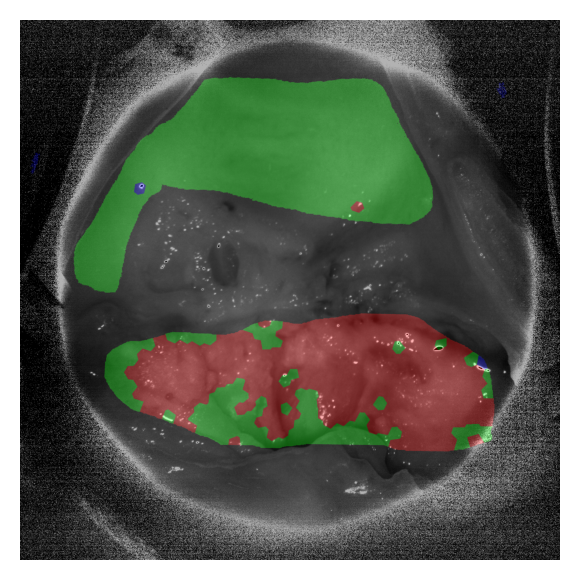}
    \end{minipage}%
    \begin{minipage}{0.14\textwidth}        
            \centering
            \includegraphics[width=\textwidth]{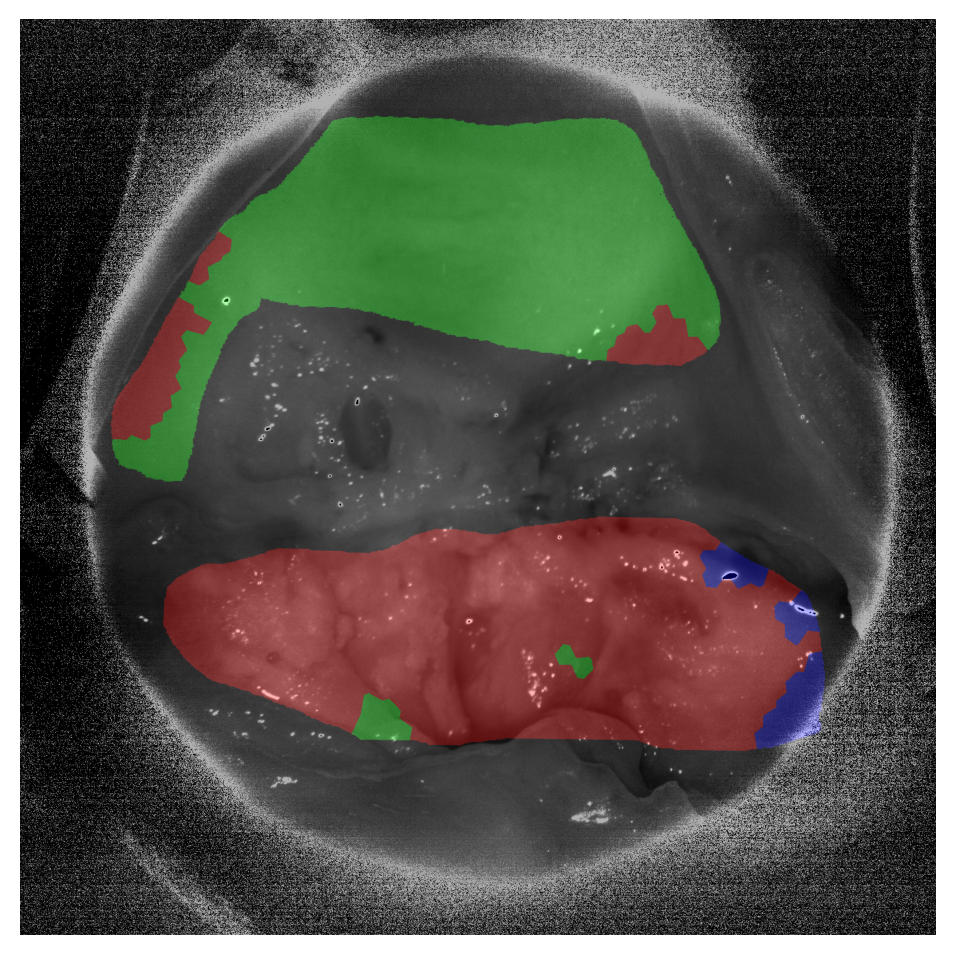}
    \end{minipage}%
    \begin{minipage}{0.14\textwidth}        
            \centering
            \includegraphics[width=\textwidth]{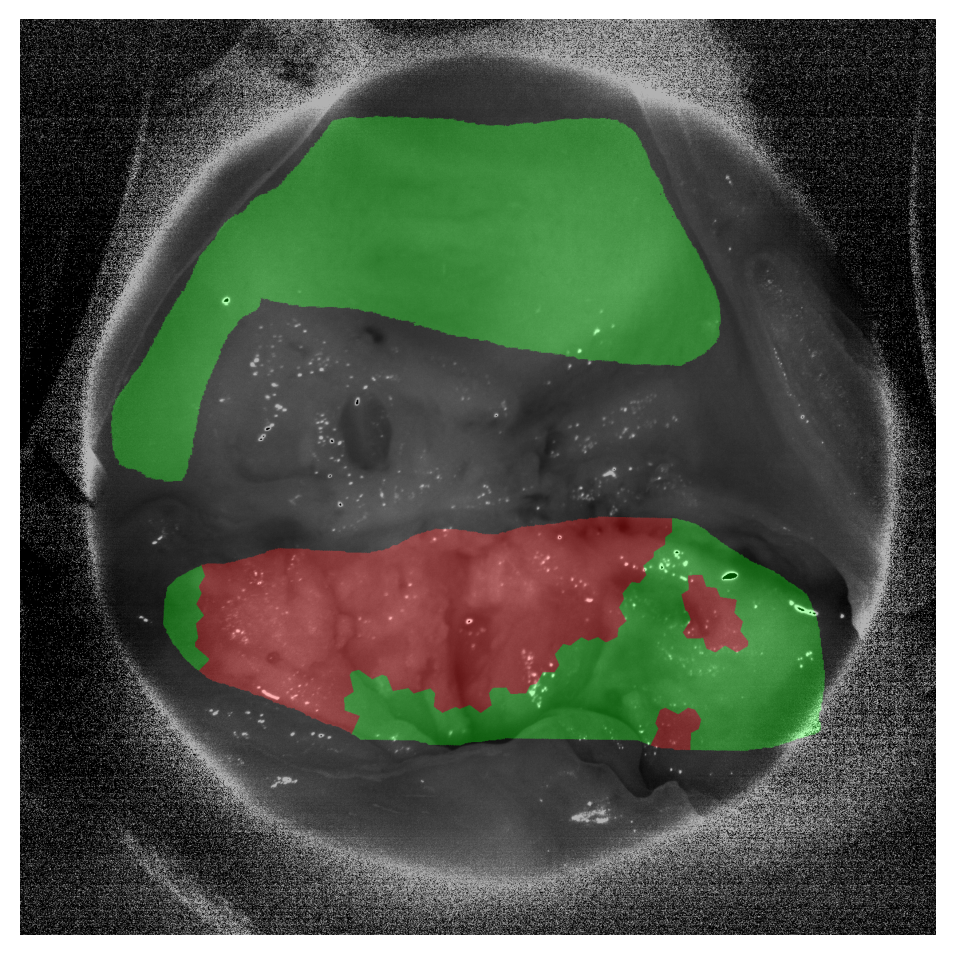}
    \end{minipage}%
    \begin{minipage}{0.14\textwidth}        
            \centering
            \includegraphics[width=\textwidth]{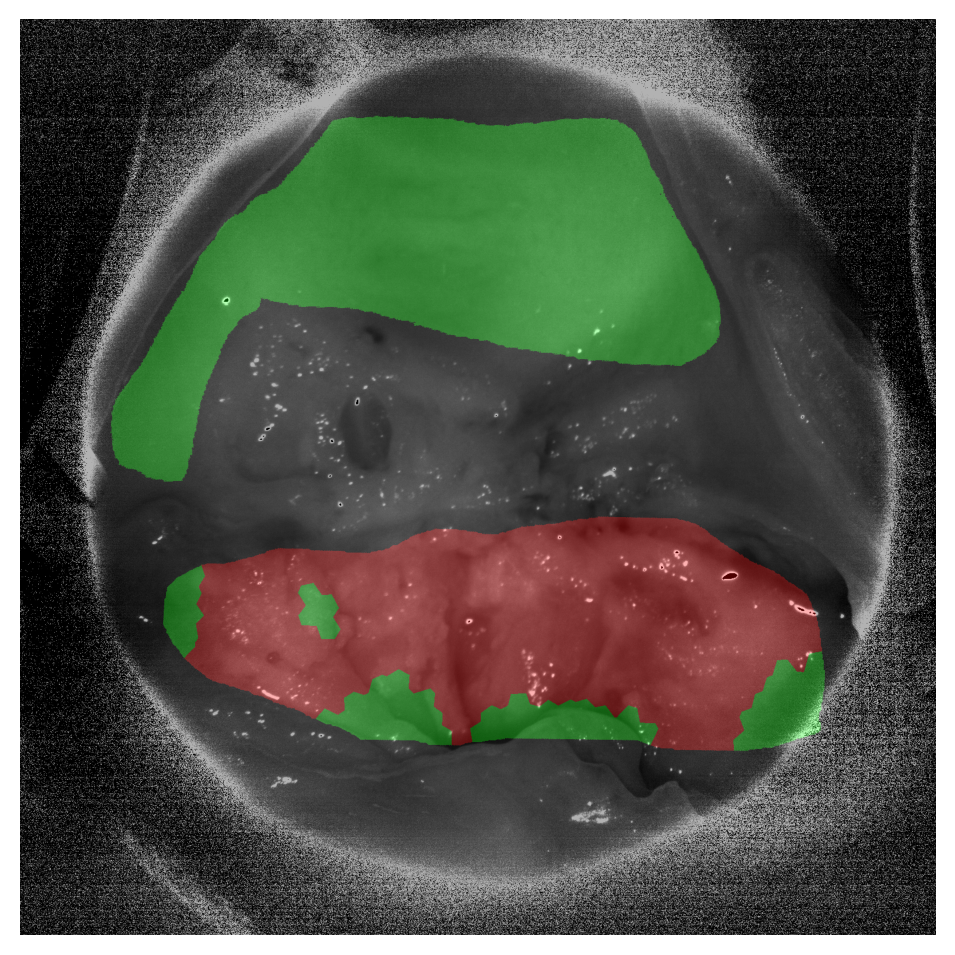}
    \end{minipage}
          \newline
    \begin{minipage}{0.14\textwidth}        
            \centering
            \includegraphics[width=\textwidth]{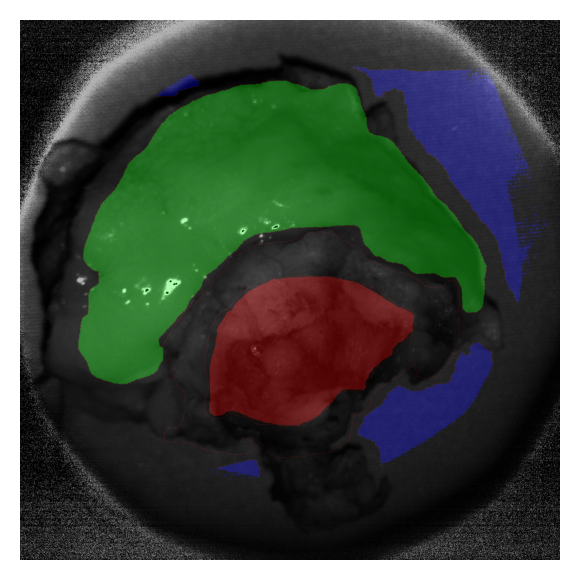}
            {\footnotesize(a) \\ GT}
    \end{minipage}%
    \begin{minipage}{0.14\textwidth}        
            \centering
            \includegraphics[width=\textwidth]{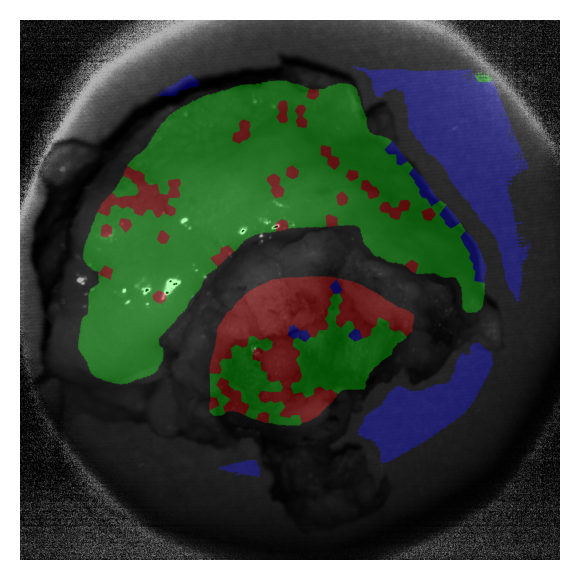}
            {\footnotesize(b) \\ CNN\_g}
    \end{minipage}%
    \begin{minipage}{0.14\textwidth}        
            \centering
            \includegraphics[width=\textwidth]{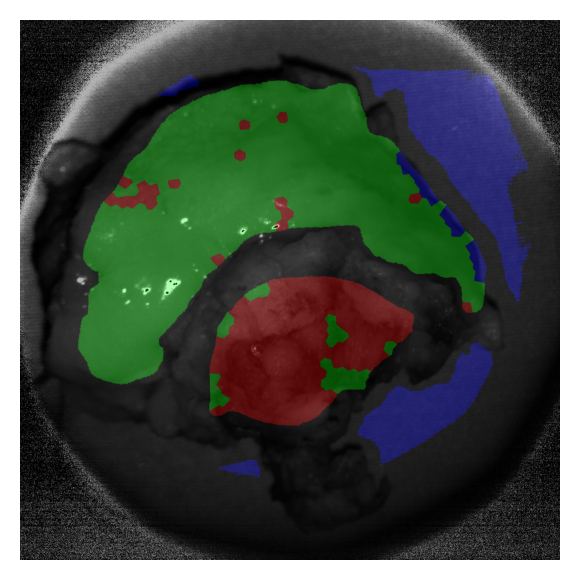}
            {\footnotesize(c) \\ CNN\_a}
    \end{minipage}%
    \begin{minipage}{0.14\textwidth}        
            \centering
            \includegraphics[width=\textwidth]{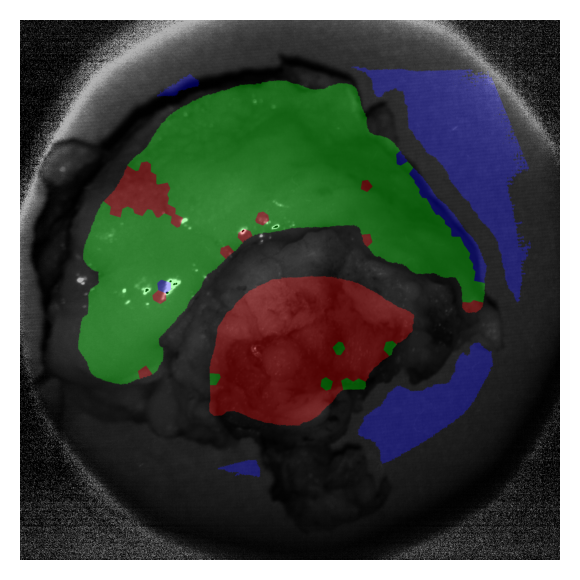}
            {\footnotesize(d)\\CNN\_aW}
    \end{minipage}%
    \begin{minipage}{0.14\textwidth}        
            \centering
            \includegraphics[width=\textwidth]{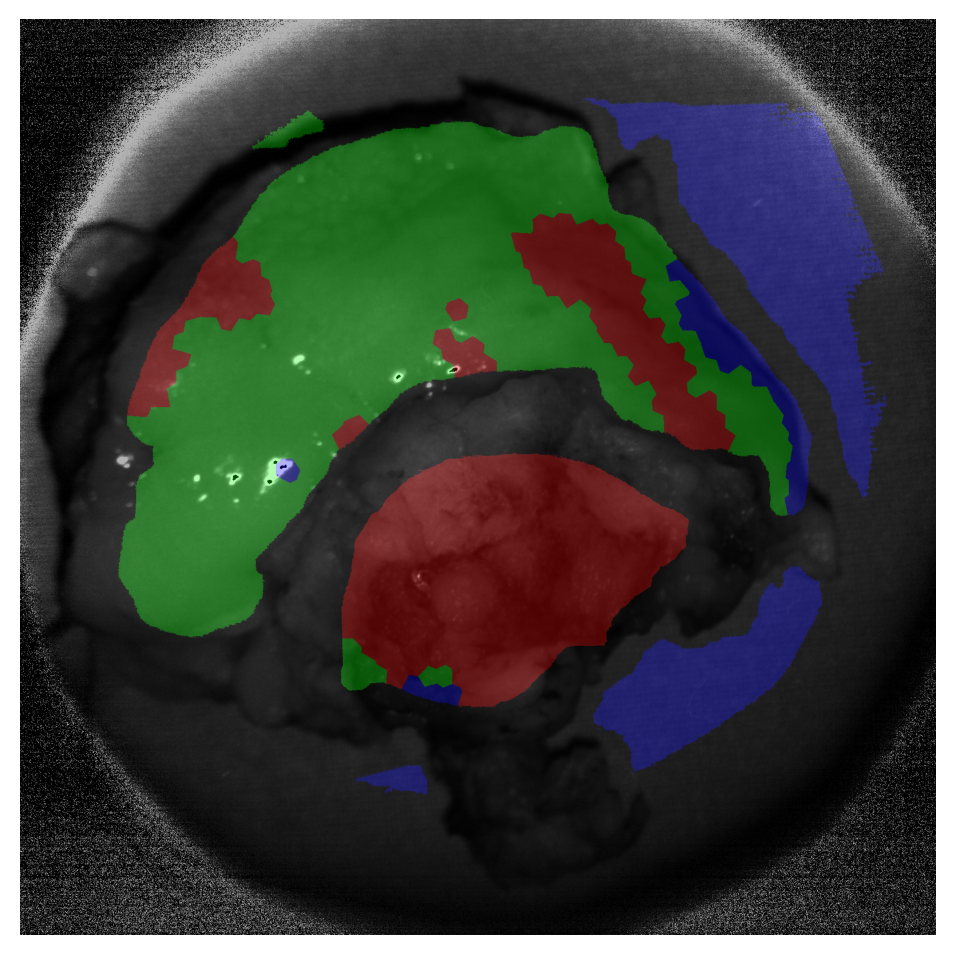}
            {\footnotesize(e) \\ GNN\_g}
    \end{minipage}%
    \begin{minipage}{0.14\textwidth}        
            \centering
            \includegraphics[width=\textwidth]{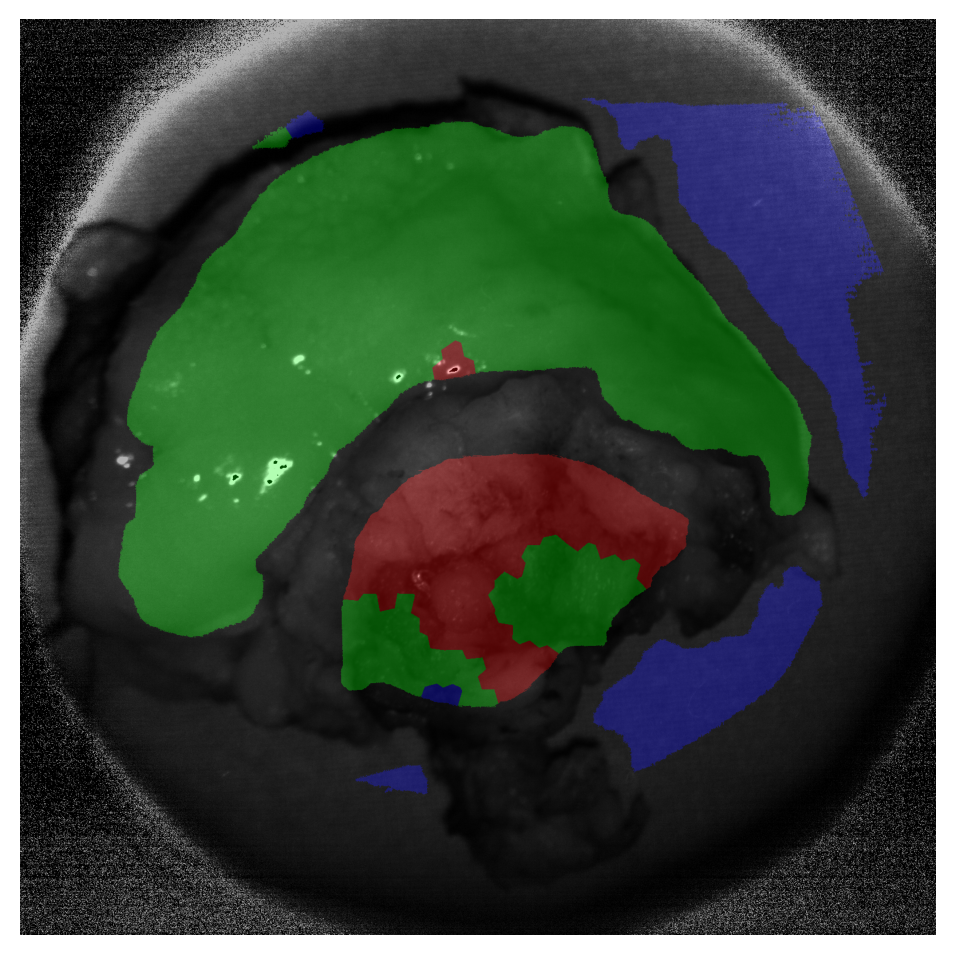}
            {\footnotesize(f) \\ GNN\_a}
    \end{minipage}%
    \begin{minipage}{0.14\textwidth}     
            \centering
            \includegraphics[width=\textwidth]{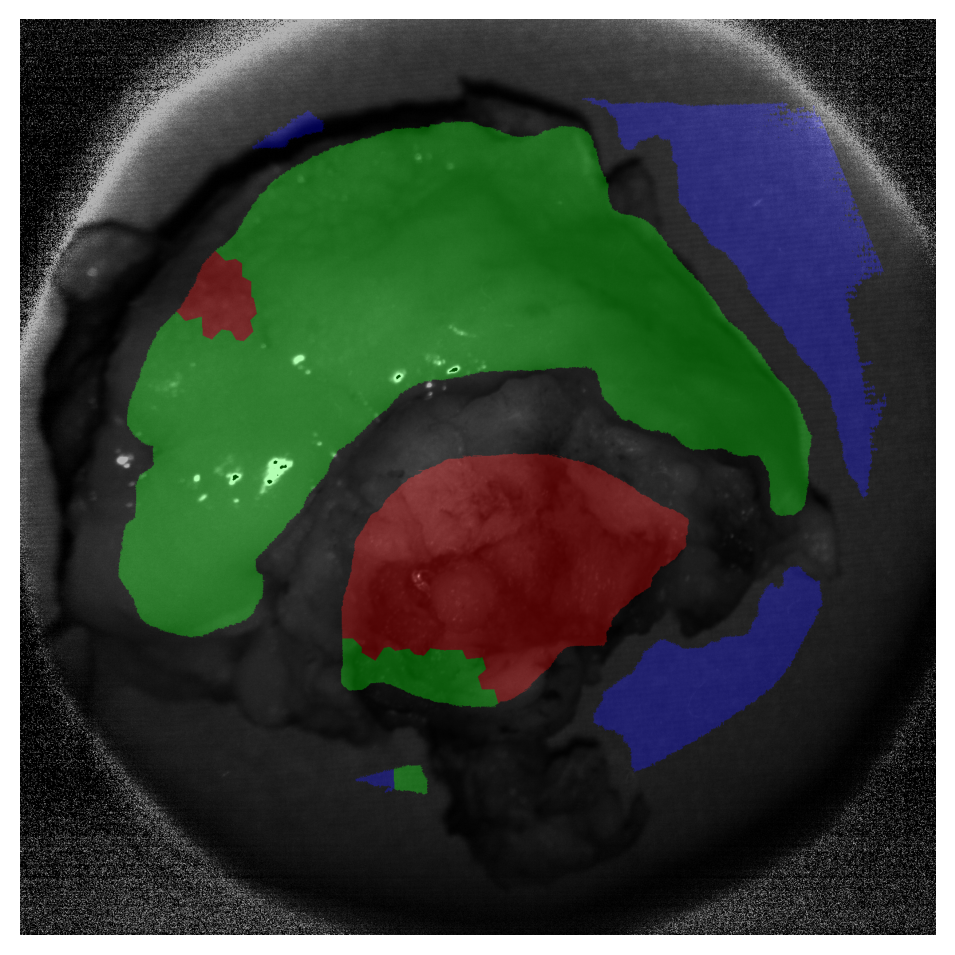}
             {\footnotesize(g) \\GNN\_aW}
    \end{minipage} 
    \newline
    \caption{
        \footnotesize Qualitative comparison of the inference results on the test set. For the different approaches described in Sec.~\ref{sec:NN_methods} and quantitatively summarized in Tab.~\ref{table:models_comparisons}.
        Each row shows the inference on one example; column (a) shows the ground truth, and the remaining columns display the results with each model.
        The grayscale images in the background represent luminosity in an intermediate wavelength channels for illustration, and 
        The overlayed colors indicate the tissue classes: red, green and blue denote tumor, healthy and background respectively.}
        \label{fig:qual_metrics}
\end{figure}

\subsection{Qualitative evaluation}

We next compare the results qualitatively on four representative images from the test set in Fig.~\ref{fig:qual_metrics}. We show for each example the ground truth followed by the segmentation predictions using each of the models of Tab.~\ref{table:models_comparisons}. 
Both CNN and GNN models succeed in producing a qualitatively correct result in most cases, such as the first three examples. The last example is a more challenging sample, where we observe a substantial fraction of false positives, which are however reduced when using more advanced approaches (columns d, g). 
For all example results, the CNN models (columns b-d) are noisier, as they cannot make use of contextual information. The GNN models (columns e-g), on the contrary, leverage context to achieve a smoother prediction.
Models trained on good tiles only (columns b, e) feature as expected a high error rate on lower-quality areas that were not seen during training, e.g., in over-/under-exposed patches and in noisy areas.
As also seen quantitatively in Tab.~\ref{table:models_comparisons}, GNN models trained on all tiles (column f) display a degraded performance on the tumor class prediction. 
A more robust prediction of this class is then restored once the weighted loss is introduced (columns d, g for CNN and GNN respectively).
The qualitative evaluation confirms that our best model, GNN\_aW, brings a significant improvement compared with the baseline from G23, CNN\_g.

\section{Conclusions and future work}
\label{sec:conclusion}

In this work, we presented a novel approach for HSI-based tumor segmentation that leverages recent advances in the field of Graph Neural Networks (GNNs) to fully exploit HSI information in both spectral and spatial domains.
We used for the first time a composite CNN+GNN model for tile-based medical HSI image segmentation. After the creation of macropixels (tiles) with an adapted version of the SLIC algorithm \cite{achanta2012slic}, we defined for each HSI image a graph object whose nodes are the tiles and whose edges connect neighboring tiles.
The node features are defined by applying a CNN feature extractor to the tile data cube, thus maximising the extraction of intra-tile spatial and spectral information. Global spatial information is in turn optimally used by passing the graph through a GAT \cite{velivckovic2017graph} graph model for node classification.

We additionally introduced a loss weighting scheme based on local image quality that aims at recovering good segmentation results also on lower-quality regions.
We evaluated this approach on a dataset of 51 HSI ex vivo images from 30 patients.
We compared the segmentation results of the GNN models with a baseline CNN by G23.
We found that all models, including the CNN baseline from G23, achieved good accuracy ($>80\%$) on a test dataset consisting of images from unseen patients.
Compared to the previous state of the art, we found that GNN models do make better use of spatial information, thus leading to more accurate and uniform results: for models trained on good-quality image regions only, GNNs improved the average accuracy with respect to the baseline from 79\% to 90\%; a qualitative analysis also confirmed the superior performance and better spatial uniformity of the GNN results.
We also demonstrated that our new quality-aware loss weighting scheme improves the robustness of the predictions for both CNN and GNN models: the average accuracy of CNN models improves from 79\% to 87\% with the weighted loss, and while the overall average accuracy of GNN models does not improve, its $F_1$-score does, as well as its qualitative spatial uniformity.

Our work shows that GNNs together with a quality-aware loss weighting strategy lead to robust tissue segmentation that generalizes to a set of 6 patients not seen during training. The next step for future research will be to apply this method to in vivo and intra-operative settings, thus paving the way for improved surgery workflow and patient outcome.

%
%
%
%
\bibliographystyle{splncs04}
\bibliography{egbib}

\end{document}